\documentclass[debug,overfull]{epl}
\usepackage{amsmath,graphicx}

\title{Self-consistent variational theory for globules}

\author{Arti Dua and Thomas A. Vilgis}
\institute{ Max-Planck-Institute for Polymer Research,
  Ackermannweg 10, 55128 Mainz, Germany
}

\begin{document}

\maketitle

\begin{abstract}
  A self-consistent variational theory for globules based on the uniform
  expansion method is presented. This method, first introduced by Edwards
  and Singh to estimate the size of a self-avoinding chain, is restricted to a
  good solvent regime, where two-body repulsion leads to chain swelling.
  We extend the variational method to a poor solvent regime where the balance
  between the two-body attractive and the three-body repulsive interactions leads to
   contraction of the chain to form a globule. By employing the  Ginzburg criterion, we
  recover the correct scaling for the $\theta$-temperature. The introduction
  of the three-body interaction term in the variational scheme recovers the
  correct scaling for the two important length scales in the globule---its
  overall size $R$, and the thermal blob size $\xi_{T}$. Since these two
  length scales follow very different statistics---Gaussian on length scales
  $\xi_{T}$, and space filling on length scale $R$---our approach extends the
  validity of the uniform expansion method to non-uniform contraction
  rendering it applicable to polymeric systems with attractive interactions.
  We present one such application by studying the Rayleigh instability of polyelectrolyte globules in poor solvents. At a critical fraction
  of charged monomers, $f_c$, along the chain backbone, we observe a clear
  indication of a first-order transition from a globular state at small $f$
  to a stretched state at large $f$; in the intermediate regime the bistable
  equilibrium between these two states shows the existence of a pearl-necklace
  structure.
\end{abstract}

\section{Introduction}

The uniform expansion method of Edwards and Singh is a well known
self-consistent variational approach to study the size and the probability
distribution of the end-to-end distance of a chain in a good solvent
\cite{edwards1,doi}. 
The method is based on the uniform expansion of a chain in terms of the
expansion in an unknown step length $b_{1}$ such that the size of the
self-avoiding chain consisting of $N$ segments is governed by Gaussian
statistics, $ {R} = N^{1/2} b_1$. The variational parameter $b_1$ is then
determined self-consistently by a perturbative calculation, usually truncated at first order. In particular, the method recovers the correct scaling,
$\left< {\bf R}^2 \right> \approx b^{2} (v/b^{3})^{2/5} N^{6/5} $, for the size of an
excluded volume chain, a result which is consistent with the classical Flory
theory \cite{flory}. In contrast to other analytical approaches---for
instance, the self-consistent field theory by Edwards \cite{edwards2}, de
Gennes \cite{degennes}, and the more rigorous renormalization group treatment
by Freed \cite{freed,freed1}---the advantage of the uniform expansion method
is its mathematical simplicity in treating the complex excluded volume
problem, which renders it applicable to more complicated systems with
additional interactions. It has become a standard method in polymer physics to
understand the equilibrium behaviour of wide variety of polymeric systems.


The uniform expansion method though applicable to wide variety of problems is
only restricted to the behavior of polymers in a good solvent regime, where
the two-body repulsive interaction swells the chain. As the temperature is
reduced from the $\Theta$ temperature to a poor solvent regime, the two-body
interaction becomes attractive; therefore, any attempt to extend this approach
to a poor solvent regime requires the stabilizing influence of the three-body
repulsive interaction. In a simple scaling form, the free energy is given by
\begin{equation}
\label{f-energy}
\beta{\cal {F}} \simeq \frac{R^{2}}{Nb^{2}} - v \frac{N^2} {R^3} + w \frac{N^3}{R^6},
\end{equation}
where $v \sim \tau b^3$ and $w \sim b^6$ are the strength of the two and three
body interactions respectively. The reduced temperature $\tau$ describes the
distance from the $\Theta$ temperature, i.e., $\tau = |T- \Theta|/ \Theta$.
Close to the $\Theta$ temperature, $v \approx 0$, and the chain is essentially
Gaussian, $R \approx N^{1/2}b$; the $\Theta$ temperature can thus be determined
by the balance of the first two terms in the above equation, which yields $v
N^2/(N^{1/2}b)^3 \approx 1$ implying $\tau \approx N^{-1/2}$. The latter is
the Ginzburg criterion which determines the temperature at which the entropic
effects (fluctuations) become less important. As the solvent quality becomes
poor, the polymer conformation is determined by the balance of the two-body
attraction and the three-body repulsion; the comparison of the two opposing
interactions (which amounts to minimizing the above free energy with respect
to $R$ and then comparing the second and the third term) gives rise to a
globular structure of size $R \approx b \tau^{-1/3}N^{1/3}$, where $b$ is the
step length. Thus the size of the globule depends not only on the number of
segments, $N$, but also on the reduced temperature, $\tau$.

Apart from the overall size of the globule there is another important length
scale in the problem---the correlation length of the density fluctuations
which defines the thermal blob size $\xi_{T}$ \cite{degennes1}. For length
scales smaller than $\xi_T$, the chain is unaffected by the attractive
interaction and essentially retains the Gaussian statistics, i.e., $\xi_{T}
\approx b {g_{T}}^{1/2}$, where $g_{T} \approx 1/\tau^2$ is the number of
monomers in the thermal blob of size $\xi_{T} \approx b/ \tau$. On length
scales larger than $\xi_{T}$ the chain begins to feel the attractive influence
of the two-body interaction and the thermal blobs are space filling, that is,
$\rho \xi_{T}^3 \approx 1$; the overall size of the globule is given by $R
\approx {\xi_{T}} {n_{T}}^{1/3}$, where $n_{T} = N/g_{T}$ is the number of
thermal blobs in a globule. In contrast to the ``uniform'' expansion in a good
solvent regime, the presence of the two different length scales in a globule
require a ``non-uniform'' contraction of the Gaussian chain. In what follows we
extend this variational method to a poor solvent regime by including the
repulsive three-body interaction along with the attractive two-body
interaction term. 

Let us first review some of the main results of the uniform expansion method,
which is essentially based on defining a new step length, $b_1 \gg b$, such
that the mean square end-to-end distance of the chain in presence of the
excluded volume is Gaussian and is governed by $\left< {\bf R}^2 \right> = N
b_1^2$. This condition amounts to replacing the original Hamiltonian ${\cal
  H}_{0} = \frac{3}{2b^2} \int _{0}^{N} ds {\dot{\bf r}}(s)^2$ with the reference
Hamiltonian ${\cal H}_{1} = \frac{3}{2b_1^2} \int _{0}^{N} ds {\dot {\bf r}}(s)^2$.
This step can easily be carried out by adding and subtracting ${\cal H}_{1}$
from ${\cal H}_{0}$ to give the following expression for the mean square
end-to-end distance:
\begin{equation}
\label{Ham}
\left< {\bf R}^2 \right> = 
\frac{\int {\cal D}[r(s)]{\bf R}^2 \exp[-({\cal H}_{1}+ 
({\cal H}_{0}-{\cal H}_{1})+{\cal H}_{2})]}{\int {\cal D}[r(s)] 
\exp[-({\cal H}_{1}+ ({\cal H}_{0}-{\cal H}_{1})+{\cal H}_{2})]},
\end{equation}
where ${\cal H}_{2} = \frac{|v|}{2}\int_{0}^{N} ds \int_{0}^{N} ds^{\prime}
\delta[{\bf r}(s)-{\bf r}(s^\prime)]$ represents the two-body interaction, and
$v = \tau b^3$ is the strength of the interaction. The idea is to expand
$\left< {\bf R}^2 \right>$ in a perturbative series about the reference
Hamiltonian ${\cal H}_{1}$ such that to the first order correction in $v$ one
obtains the following variational equation for the unknown parameter $b_1$:
\begin{equation}
\label{eqn1}
\left< {\bf R}^2 \right>_{1} \left< {\cal H}_{0}-{\cal H}_{1}+{\cal H}_2 \right>_{1} - \left< {\bf R}^2 ({\cal H}_{0}-{\cal H}_{1}+{\cal H}_2) \right>_{1} = 0,
\end{equation}
where $\left< \cdots \right>$ denotes the average with respect to the
probability distribution 
$$
{\cal P}[{\bf r}(s)] \propto \exp \left[{-\frac{3}{2b_1^2} \int _{0}^{N} ds
    {\dot{\bf r}}(s)^2}\right].
$$  
The Gaussian nature of the probability
distribution makes the evaluation of the above averages simple. The
details of the calculation can be found in References (1) and (2); here we
simply present the result:
\begin{equation}
\label{eqn2}
\alpha^{5} - \alpha^3 - 2 v \left({\frac{6}{\pi^3}}\right)^{1/2}\frac{N^{1/2}}{b^3} = 0,
\end{equation}
where $\alpha = R/R_{0}$ is the expansion factor and $R_{0} = N^{1/2}b$. For a
swollen chain, $\alpha \gg 1$, the above equation can be solved to give $R
\approx |v|^{1/5}b^{2/5}N^{3/5}$, which recovers the correct scaling for the
excluded volume chain. It is interesting to note that inclusion of the higher
order terms only changes the numerical factors without altering the overall
structure of $R$ \cite{edwards1,doi}.

In presence of the attractive two-body interaction term Eq. (\ref{eqn1}) has
to be modified to include the effects of the three-body repulsive interaction.
The resulting expression is given by
\begin{equation}
\label{eqn3}
\left< {\bf R}^2 \right>_{1} \left< {\cal H}_{0}-{\cal H}_{1}-{\cal H}_2 + {\cal H}_3\right>_{1} - \left< {\bf R}^2 ({\cal H}_{0}-{\cal H}_{1} - {\cal H}_2 + {\cal H}_3) \right>_{1} = 0.
\end{equation}
where $w = b^6$ is the strength of the three-body interaction and 
$${\cal
  H}_{3} = \frac{w}{6}\int_{0}^{N} ds \int_{0}^{N} ds^{\prime} \int_{0}^{N}
ds^{\prime\prime} \delta[{\bf r}(s)-{\bf r}(s^\prime)] \delta[{\bf
  r}(s^\prime)-{\bf r}(s^{\prime\prime})].$$ 
The quantity we are interested in
evaluating is $\left< {\bf R}^2 \right>_{1}\left< {\cal H}_3\right>_{1} -
\left< {\bf R}^2 {\cal H}_3 \right>_{1}$, where, as before, the average with
respect to the Gaussian distribution function can easily be performed to give
\begin{eqnarray}
\label{3body1}
\left< {\bf R}^2 \right>_{1}\left< {\cal H}_3\right>_{1} - \left< {\bf R}^2 {\cal H}_3 \right>_{1} &=& \frac{w b_{1}^4}{9 (2\pi)^6}\int_0^{\infty} dq\int_{0}^{\infty} dk  \int_{0}^{N} ds \int_{0}^{s} ds^{\prime} \int_{0}^{s^\prime} ds^{\prime\prime} ~~[(s-s^\prime)^2 k^4 q^2 \nonumber\\
& & ~~~~~~~~~~~~~+ (s^{\prime}-s^{\prime\prime})^2 k^2 q^4]e^{-k^2 b_{1}^2(s-s^\prime)/6} e^{-q^2 b_{1}^2(s^{\prime}-s^{\prime\prime})/6}. 
\end{eqnarray} 
It is clear from the structure of the above expression that the integration
over $k$ and $q$ followed by integrations over the internal coordinates of the
chain (e.g., $s^{\prime\prime}$, $s^\prime$ and $s$) leads to a divergence. This
divergence is a common feature of excluded-volume problems in polymer physics,
and has it origins in the use of delta function pseudopotentials. This
divergence, though present even in the case of the two-body interaction, cancels
in the overall expression for $\left<{\bf R}^2\right>$ (see References (1) and
(2) for more details). This is not the case when the three-body interaction is
included, and the evaluation of above integrations turns out to be
non-trivial. However, at the present level of the mean-field description,
it is possible to make reasonable approximations to probe the relevant length
scales in a globule.

The idea is to first evaluate the $k$ integration in the first term and the
$q$ integration in the second term of Eq. (\ref{3body1}) followed by
integrations over the internal coordinates of the chain ($s^{\prime\prime}$,
$s^\prime$ and $s$). This order of integration is important to be able to
perform all but the last integration without encountering any divergence. The
result of such an operation can be expressed as
\begin{equation}
\label{3body2}
\left< {\bf R}^2 \right>_{1}\left< {\cal H}_3\right>_{1} - \left< {\bf R}^2 {\cal H}_3 \right>_{1} = \frac{w N}{ {b_{1}}^4}\left(\frac{81}{(2\pi)^6}\right) \left[ \left(\frac{2}{3}\right)^{3/2}\int_{0}^{\infty}dq ~~N^{1/2}b_{1} - \frac{\pi^{3/2}}{2} \right],
\end{equation}
which diverges as $q \rightarrow \infty$. However, within the mean-field
description, the divergence can be removed by introducing an upper cut-off for
the wave number $q$. This is not a serious approximation since density
fluctuations in a globule are extremely small, and the mean-field description
suffices. Thus by introducing an upper cut-off, $q_{max} = 2\pi/\Lambda$, one
can probe the relevant length scales, $\Lambda$, in a globule. The result of
the above integration when substituted into Eq. (\ref{eqn3}), produces the
following equation for $\alpha$:
\begin{equation}
\label{eqn4}        
\alpha^{5} - \alpha^3 + \tau N^{1/2} - \frac{N^{1/2}b}{ \alpha^2 \Lambda}  = 0,
\end{equation}
where we have used $v = \tau b^3$ and $w = b^6$. For convenience, we have
ignored all the numerical coefficients.  First, we begin by asking for the
solution of the above equation for length scale of the size of the chain,
$\Lambda = R$. Since the only known length scale for the average size of the
chain is $R = N^{1/2} b_{1}$, it is reasonable to substitute $\Lambda \approx
N^{1/2}b_{1}$ into the above equation to yield
\begin{equation}
\label{eqn5}
\alpha^{5} - \alpha^3 + \tau N^{1/2} - \frac{1}{ \alpha^3}= 0.   
\end{equation}
In presence of the attractive potential, the chain adopts a globular
conformation, and entropic effects become less important. We can, therefore,
use the Ginzburg criterion to estimate the $\Theta$ temperature by comparing
the first term (containing effects due to entropy) with the second term
(containing effects due to the attractive potential) in the above equation for
the condition $\alpha \ll 1$.  Such a comparison yields $\tau \approx
N^{-1/2}$, a result which is consistent with the scaling analysis. The balance
between the second and the third term, on the other hand, yields the size of
the globule, $R = b\tau^{-1/3}N^{1/3}$.

Second, for length scales smaller than the size of the globule but larger than the step length $b$, we look for the density correlations of the order of the size of the thermal blob, i.e., $\Lambda \approx \xi_{T}$. The substitution of the latter expression into Eq. (\ref{eqn4}) yields
\begin{equation}
\label{eqn5}
\alpha^{5} - \alpha^3 + \tau N^{1/2} - \frac{N^{1/2}b}{ \alpha^2 \xi_{T}}= 0. 
\end{equation}
Since the statistics inside a thermal blob is Gaussian, we determine $\xi_{T}$
for the condition $\alpha = 1$. The latter condition produces $\xi_{T} =
b/\tau$, which, as discussed in the Introduction, represents the size of the
thermal blob. Moreover, the Gaussian statistics for $\xi_{T}$ implies $\xi_{T}
= {g_{T}}^{1/2} b$, where $g_{T}$ is the number of monomers inside the thermal
blob. The latter result along with $\xi_{T} = b/\tau$ leads to $g_{T} =
1/\tau^2$. When the size of the thermal blob $\xi_{T}$ is of the order of the
step length $b$, we recover the condition, $\tau \approx 1$, defining the most
compact globule of size given by $R \approx b N^{1/3}$.

In contrast to the scaling theories, the present approach provides an alternative way to estimate the blob size. In presence of the  repulsive three-body interaction, the resulting variational equation needs to be regularized, a requirement that leads to a nonuniform contraction of the chain. The presence of the three body interaction is, therefore, essential to compute the overall size as well as the internal structure of the globule. 

Thus our extension of the method of Edwards and Singh to the nonuniform case recovers all the essential features of a globular structure, and yields a simple variational scheme which can be applied to systems with attractive interactions. In the next section we use this approach to study the Rayleigh instability of charged globules in poor solvents.

\section{Polyelectrolytes in poor solvents}

In addition to the two-body attractive and three-body repulsive interactions,
the conformation of polyelectrolytes in poor solvents is also governed by the
long-range electrostatic interaction between charged monomers. With the
increase in the fraction of charges, $f$, along the chain backbone, a conformational transition occurs from a compact globular state at small $f$
to a highly extended rod-like state at large $f$. This transition is due the
repulsive nature of the Coloumb interaction, which at a critical fraction of
charges makes the large globular structure highly unstable; the system gains
energy by splitting into smaller charged globules connected by a narrow string
to form what is commonly known as a pearl-necklace---a string of locally
collapsed pearls (globules) \cite{avd,thomas}. This instability of a charged globule
is believed to be similar to the Rayleigh instability of a charged liquid
droplet which occurs as soon as its electrostatic energy, $Q^2k_{B}Tl_{B}/e^2
R$, exceeds the surface energy, $\gamma R^2$, i.e., $Q > (\gamma R^3/k_{B}T
l_{B})^{1/2} e$ \cite{kantor1,kantor2}, where $Q$ is the charge on the droplet
of size $R$, $e$ is the elementary charge, $l_{B} = e^2/4\pi\epsilon k_{B}T$
is the Bjerrum length, $\epsilon$ is the dielectric constant of the solvent
and $\gamma$ is the surface tension. As a result of this instability a charged
droplet spontaneously breaks up into smaller droplets, placed infinitely apart, each with a charge less than the critical one. However, due
to the constraint of chain connectivity the preferred equilibrium conformation
of a charged globule is that of a pearl-necklace, which with the increase in
$f$ results in a stretched state. Although scaling theories do provide some
evidence of the abrupt transition at a critical fraction of the charge from a
globule to a stretched state, a clear indication of its being first-order is
still missing. In what follows we study in detail the nature of the transition
that brings about the Rayleigh instability in a charged globule; the approach
we use is the variational method discussed in the last section.

The repulsion between charged monomer is described by the Coulomb interaction
given by ${\cal H}_{c} = \frac{v_{c}}{2} \int_{0}^{N} ds \int_{0}^{N}
ds^{\prime} |{\bf r}(s) - {\bf r}(s^{\prime})|^{-1}$, where $v_{c} = f^2
l_{B}$ is the strength of the Coloumb interaction. This electrostatic
interaction when included into Eqs. (\ref{Ham}) and (\ref{eqn3}) yields the
following equation:
\begin{equation}
\label{eqn-c}
\left< {\bf R}^2 \right>_{1} \left< {\cal H}_{0}-{\cal H}_{1}-{\cal H}_2 + {\cal H}_3 + {\cal H}_c\right>_{1} - \left< {\bf R}^2 ({\cal H}_{0}-{\cal H}_{1} - {\cal H}_2 + {\cal H}_3 + {\cal H}_c) \right>_{1} = 0.
\end{equation}
The additional term that needs to be calculated is given by $\left< {\bf R}^2 \right>_{1}\left< {\cal H}_c\right>_{1} - \left< {\bf R}^2 {\cal H}_c \right>_{1}$. The details of the calculation will be presented elsewhere \cite{a-t}; here we simply present the result:
\begin{equation}   
\label{eqn-c1}
\left< {\bf R}^2 \right>_{1}\left< {\cal H}_c\right>_{1} - \left< {\bf R}^2 {\cal H}_c \right>_{1} = - \frac{4}{45}\left(\frac{6}{\pi}\right)^{1/2} f^2 l_{B}N^{5/2}b_{1}
\end{equation}
\begin{figure}
\rotatebox{270}{\includegraphics[width=8cm,height=14cm]{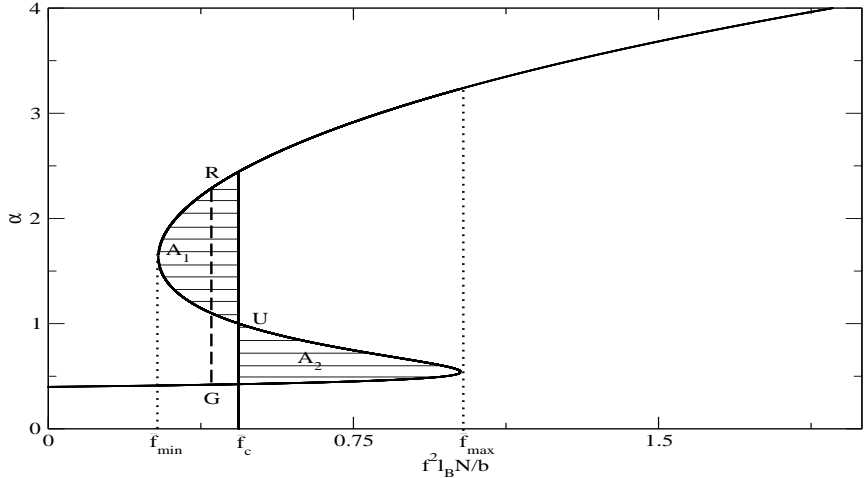}}
\caption{ The numerical solution of Eq. (\ref{eqn-c2}) for $\alpha$ is plotted
  as a function of the strength of the electrostatic energy ${\bar f} = f^2 l_{B}N/b$ for $\tau = 0.5$ and $N = 1000$. The region between the dotted lines is the
  region of coexistence between a globule and an extended state. The bistable
  equilibrium between these two states is represented by the dashed line,
  where $R$ indicates the rod-like extended state and $G$ indicates the
  globular state; the unstable part of the region is represented by the curve
  $U$. ${\bar f}_c$ represents the critical fraction of charge at which the
  first-order transition from a globular to an extended state occurs. The
  parallel horizontal lines represent regions $A_1$ and $A_2$ of equal area.}
\end{figure}
The above expression when substituted into Eq. (\ref{eqn-c}) gives the
following equation for $\alpha$:
\begin{equation}   
\label{eqn-c2}
\alpha^{5} - \alpha^3 + \tau N^{1/2} - \frac{1}{ \alpha^3} - \left(\frac{f^2 l_{B}}{b}\right)N^{3/2}\alpha^2= 0. 
\end{equation}
The above equation for $\alpha$ can be solved in the following two limiting cases: when
$f^2 l_{B}N/b \ll \tau$, the attractive influence of the two-body interaction
dominates, and the comparison of the third and the fourth term yields a
globule of size $R = b \tau^{-1/3} N^{1/3}$; in the opposite limit of $f^2
l_{B}N/b \gg \tau$, the repulsive effects of the electrostatic interaction
overcomes the attractive interaction, and the comparison of the first and the
last term yields $\alpha = (f^2l_{B}/b)^{1/3} N^{1/2}$ implying a stretched
state, $R = bN(f^2l_{B}/b)^{1/3}$. In terms of the electrostatic blob picture
the latter represents the linear arrangement of $N/g_{el}$ (electrostatic)
blobs of size $\xi_{el} = (f^2 l_{B}/b)^{-1/3}$, where $g_{el} =
(l_{B}f^2/b)^{-2/3}$ is the number of monomers in a blob. Therefore, the
expressions obtained from the variational approach are in complete agreement
with the scaling predictions \cite{avd,thomas}.
      
In the intermediate regime, ${\bar f} = f^2 l_{B}N/b \simeq \tau$, Eq. (\ref{eqn-c2}) can only be solved numerically. The result is presented in the form of Fig. (1),
which is a plot of $\alpha$ as a function of the strength of the electrostatic
energy. In the intermediate regime (the region between the dotted lines
represented by ${\bar f}_{min}$ and ${\bar f}_{max}$) the curve folds back on itself to give a bistable equilibrium between a globular state ($\alpha < 1$) and a
rod-like state ($\alpha > 1$). The coexistence of these two states in the
region between the two dotted lines in Fig.1 is a clear indication of the
existence of the pearl-necklace structure in this region. Moreover, the abrupt
transition from a globular to an extended state is evidently first-order in
nature. Therefore, we use the Maxwell equal-area construction to determine the
critical fraction of charge ${\bar f}_{c}$ at which the spontaneous Rayleigh
splitting of the charged globule occurs. The point ${\bar f}_{c}$ in Fig. 1 is the point which ensures that the area under the two curves (represented by
$A_1$ and $A_2$) is equal. The intermediate regime corresponds to the balance between the forces close to the Rayleigh splitting, i.e., when the electrostatic force $k_{B}Tl_{B}f^{2}N^{2}/R^{2}_{pn}$ becomes equal to the line tension $k_{B}T\tau/b$. The chain extension then scales as $R_{pn} = (l_{B}f^{2}/b\tau)^{1/2} N b$, where $R_{pn}$ represents the mean size of the pearl-necklace structure.

The pearl-necklace structure, as seen in simulations, is a highly fluctuating
structure both in terms of the position and the size of pearls \cite{limbach}.
In the mean-field description, the existence of such a structure is evident
through a bistable equilibrium between a globular and a stretched state, as is
clear from the coexistence region, ${\bar f}_{min} \leq {\bar f} \leq {\bar f}_{max}$, in Fig. 1. The scaling theories usually estimates the average size of a pearl-necklace by comparing its surface energy with that of the electrostatic energy. In a
future publication, we will present a detailed analysis of the dependence of
$R$, in the coexistence region, on $f$, $N$ and $\tau$ to make comparison with
the scaling predictions. The effects of the variation of $\tau$ and $N$ on the
coexistence region (pertaining to pearl-necklaces) and $f_c$ will also be
discussed \cite{a-t}.

\section{Conclusions}
      
We have extended the uniform expansion method of Edwards and Singh to a poor solvent regime by accounting for the three-body repulsive interaction in addition to the two-body attractive potential. The self-consistent variational approach recovers the two important length scales in a globule---its overall size and the thermal blob size. This method can be applied to several different polymeric systems where the effective two-body potential is negative. To illustrate the generality of our approach, we applied this method to polyelectrolytes in poor solvents to study the Rayleigh instability of charged globules in detail. The presence of the additional long-range repulsion due to the electrostatic interaction between the charged monomers destabilizes the globular structure as the fraction of charged monomers along the chain backbone is increased. At a certain critical fraction of charges the globule splits into locally collapsed monomers joined by a linear string to form a pearl-necklace. Using the present approach, we observe a clear indication of a first-order transition, at a critical fraction of charged monomers, from a globular to an extended state. The intermediate regime of bistable equilibrium between these two states is the region where the pearl-necklace structure is stable.

In contrast to flexible polymers, the behaviour of semiflexible polymers in poor solvents is not well understood. Given the simplicity of the present approach, it holds out the possibility of addressing many such issues, including the equilibrium behaviour of biopolymers with specific interactions \cite{a-t}.

\end{document}